\pdfoutput=1

\documentclass[aps,prd,twocolumn,superscriptaddress,showpacs]{revtex4}
\usepackage{graphics}
\usepackage{epstopdf}
\usepackage{graphicx}
\usepackage{dcolumn}
\usepackage{bm}
\usepackage{amsmath}
\usepackage{color}

\newcommand{\be}{\begin{equation}}
\newcommand{\ee}{\end{equation}}
\newcommand{\ba}{\begin{eqnarray}}
\newcommand{\ea}{\end{eqnarray}}
\newcommand{\bfig}{\begin{figure}[t]\begin{centering}}
\newcommand{\efig}{\end{centering}\end{figure}}

\def\fer{{\it Fermi~}}

\begin{document}

\title{Gamma Ray Bursts in the Swift-Fermi Era}

\author{Neil Gehrels}
\email{neil.gehrels@nasa.gov}
\affiliation{Astrophysics Science Division, 
NASA Goddard Space Flight Center, Greenbelt, MD 20771 USA}

\author{Soebur Razzaque}
\email{srazzaqu@gmu.edu}
\affiliation{George Mason University, Fairfax, VA 22030, USA; 
Resident at Space Science Division, Naval Research Laboratory, 
Washington, DC 20375 USA}

\date{\today}

\begin{abstract}
Gamma-ray bursts (GRBs) are among the most violent occurrences in the
universe.  They are powerful explosions, visible to high redshift, and
thought to be the signature of black hole birth. They are highly
luminous events and provide excellent probes of the distant universe.
GRB research has greatly advanced over the past 10 years with the
results from Swift, Fermi and an active follow-up community.  In this
review we survey the interplay between these recent observations and
the theoretical models of the prompt GRB emission and the subsequent
afterglows.
\end{abstract}                            
 
\maketitle

\tableofcontents

\section{Introduction}
\label{sec:intro}

GRBs are the most extreme explosive events in the Universe.  The
initial prompt phase lasts typically less than 100 s and has an energy
content of $\sim 10^{51}$ ergs, giving a luminosity that is a million
times larger than the peak electromagnetic luminosity of a supernova.
The GRB name is a good one because their spectra peak in the gamma-ray
band between $\sim 100$ keV and $\sim 1$ MeV.  The source of the
energy powering the bursts is thought to be the gravitational collapse
of matter to form a black hole or other compact object.

GRBs were discovered in the late 1960s by the Vela satellites
monitoring the Nuclear Test Ban Treaty between the US and the Soviet
Union \cite{Klebesadel+73}.  It was slow progress for 20 years
learning the origin of these brilliant flashes.  Gamma-ray instruments
of the time had poor positioning capability so only wide-field,
insensitive telescopes could follow-up the bursts to look for
counterparts at other wavelengths.  Nothing associated with the GRBs
was seen in searches hours to days after their occurrence.  In the
1990's the Burst and Transient Source Experiment (BATSE) onboard the
Compton Gamma Ray Observation (CGRO) obtained the positions of $\sim
3000$ GRBs and showed that they were uniformly distributed on the
sky \cite{Meegan+92}, indicating either an extragalactic or
galactic-halo origin.  BATSE also found that GRBs separate into two
duration classes, short and long GRBs, with dividing line at $\sim 2$
s \cite{Kouveliotou+93}.  The detection of X-ray afterglows and more
accurate localizations delivered by the BeppoSAX
mission \cite{Costa+97,Paradijs+97} enabled optical redshifts of GRBs
to be measured and their extragalactic origin to be confirmed.  The
long bursts were found to be associated with far-away galaxies at
typical redshifts $z=1-2$, implying energy releases in excess of
$10^{50}$ ergs, after taking into account the jet beaming factor.  The
afterglow time decay often steepened after a day indicating a
geometrical beaming of the radiation into a jet of opening angle $\sim
5^\circ$ \cite{Harrison+99}.

Key open issues prior to the Swift and Fermi era included: (1) the
origin of short GRBs, (2) the nature of the high energy radiation from
GRBs, (3) the redshift distribution of bursts and their usage for
early universe studies, and (4) the physics of the jetted outflows.

Swift and Fermi, launched in 2004 and 2008 respectively, have opened a
new era in GRB research.  

Swift is a NASA mission and Fermi a NASA collaboration with the US
Department of Energy.  They are both major international partnerships
and have different and complementary capabilities.  Swift has a
wide-field imaging camera in the hard X-ray band that detects the
bursts at a rate of $\sim 90$ per year, providing positions with
arcminute accuracy.  The spacecraft then autonomously and rapidly (100
s) reorients itself for sensitive X-ray and UV/optical observations of
the afterglow.  Fermi has two wide-field instruments.  One detects
bursts in the gamma-ray band at a rate of $\sim 300$ per year,
providing spectroscopy and positions with 10-degree accuracy.  The
other observes bursts in the largely-unexplored high-energy gamma-ray
band at a rate of $\sim 10$ per year.  Combined, the two missions are
advancing our understanding of all aspects of GRBs, including the
origin of short bursts, the nature of bursts coming from the explosion
of early stars in the universe and the physics of the fireball
outflows that produce the gamma-ray emission.

\section{Observations}
\label{sec:obs}

\subsection{Swift GRBs}
\label{subsec:swift}

The Swift mission \cite{Gehrels+04} has three instruments: the Burst
Alert Telescope (BAT; \cite{Barthelmy+05}), the X-Ray Telescope
(XRT; \cite{Burrows+05}) and the UV Optical Telescope
(UVOT; \cite{Roming+05}).  The BAT detects bursts and locates them to
$\sim 2$ arcminute accuracy.  The position is then sent to the
spacecraft to repoint the XRT and UVOT at the event.  Positions are
also rapidly sent to the ground so that ground telescopes can follow
the afterglows.  There are more than 50 such telescopes of all sizes
that participate in these world-wide follow-up campaigns.
Measurements of the redshift and studies of host galaxies are
typically done with large ground-based telescopes, which receive
immediate alerts from the spacecraft when GRBs are detected.  Swift
has, by far, detected the largest number of well-localized bursts with
afterglow observations and redshift determinations.  As of 1 September
2012, BAT has detected 709 GRBs (annual average rate of $\sim 90$ per
year).  Approximately $80\%$ of the BAT-detected GRBs have rapid
repointings (the remaining $20\%$ have spacecraft constraints that
prevent rapid slewing).  Of those, virtually all long bursts observed
promptly have detected X-ray afterglow.  Short bursts are more likely
to have negligible X-ray afterglow, fading rapidly below the XRT
sensitivity limit.  The fraction of rapid-pointing GRBs that have UVOT
detection is $\sim 35\%$.  Combined with ground-based optical
observations, about $\sim 60\%$ of Swift GRB have optical afterglow
detection.  There are 225 Swift GRBs with redshifts compared with 41
pre-Swift.

\begin{figure}
\includegraphics[width=3.2in]{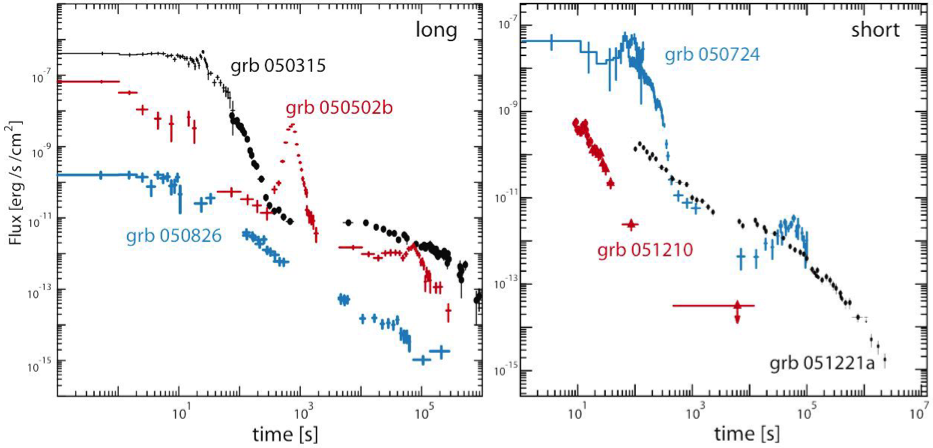}
\caption{
Representative examples of X-ray afterglows of (a) long and (b) short
Swift events with steep-to-shallow transitions (GRB050315, 050724),
large X-ray flares (GRB050502B, 050724), and rapidly declining
(GRB051210) and gradually declining (GRB051221a, 050826; flux scale
divided by 100 for clarity) afterglows. From Gehrels, Ramirez-Ruiz and
Fox (2009) \cite{Gehrels+09}.  }
\label{fig:SwiftXray}
\end{figure}

Swift was specifically designed to investigate GRB afterglows by
filling the temporal gap between observations of the prompt emission
and the afterglow \cite{OBrien+06}. Figure \ref{fig:SwiftXray} shows
example X-ray afterglows from XRT from long and short
GRBs \cite{Gehrels+09}. The combined power of the BAT and XRT has
revealed that in long GRBs the non-thermal prompt X-ray emission
smoothly transitions into the decaying afterglow.  Often, a
steep-to-shallow transition is found suggesting that prompt emission
and the afterglow are distinct emission components \cite{Nousek+06,
Zhang+06}.  The early steep-decay phase seen in the majority of GRBs
is a surprise.  The current best explanation is that we are seeing
high-latitude emission due to termination of central engine
activity \cite{Zhang+06}.

Swift has discovered erratic flaring behavior
(Figure \ref{fig:SwiftXray}), lasting long after the prompt phase in
$\sim 30\%$ of X-ray afterglows.  The most extreme examples are flares
with integrated power similar to or exceeding the initial
burst \cite{Burrows+05b}.  The rapid rise and decay, multiple flares in
the same burst, and cases of fluence comparable with the prompt
emission, suggest that these flares are due to the central engine.

Long GRBs (LGRBs) are associated with the brightest regions of
galaxies where the most massive stars occur \cite{Fruchter+06}.  LGRBs
occur over a large redshift range from $z = 0.0085$ (GRB 980425) to
$z>8$ (GRB 090423 \& GRB 090429B).  With few exceptions, LGRBs that
occur near enough for supernova detection have accompanying Type Ib or
Ic supernovae.  The exceptions are cases of bursts that may be
misclassified as long or that may have exceptionally weak supernovae.
These facts support the growing evidence that long bursts are caused
by ``collapsars'' where the central core of a massive star collapses
to a compact object such as a black hole \cite{MacFayden+99} or
possibly a magnetar \cite{Usov92, Soderberg+06}.

Theories for the origin of LGRBs predate Swift, but are supported by
the new data.  In them, a solar rest mass worth of gravitational
energy is released in a very short time (seconds or less) in a small
region of the order of tens of kilometers by a cataclysmic stellar
event. The energy source is the collapse of the core of a massive
star. Only a small fraction of this energy is converted into
electromagnetic radiation, through the dissipation of the kinetic
energy of a collimated relativistic outflow, a fireball with bulk
Lorentz factors of $\Gamma\sim 300$, expanding out from the central
engine powered by the gravitational accretion of surrounding matter
into the collapsed core or black hole.

\subsection{Fermi GRBs}
\label{subsec:fermi}

Mission \& Statistics --- The Fermi instruments are the Gamma-ray
Burst Monitor (GBM, \cite{Meegan+09}) and the Large Area Telescope
(LAT, \cite{Atwood+09}).  The GBM has scintillation detectors and
covers the energy range from 8 keV to 40 MeV.  It measures spectra of
GRBs and determines their position to $\sim 5^\circ$ accuracy.  The
LAT is a pair conversion telescope covering the energy range from 20
MeV to 300 GeV.  It measures spectra of sources and positions them to
an accuracy of $<1^\circ$.  The GBM detects GRBs at a rate of {$\sim
300$} per year, of which on average $20\%$ are short bursts.  The LAT
detects bursts at a rate of {$\sim 10$} per year.

\begin{figure}
\includegraphics[width=3.2in]{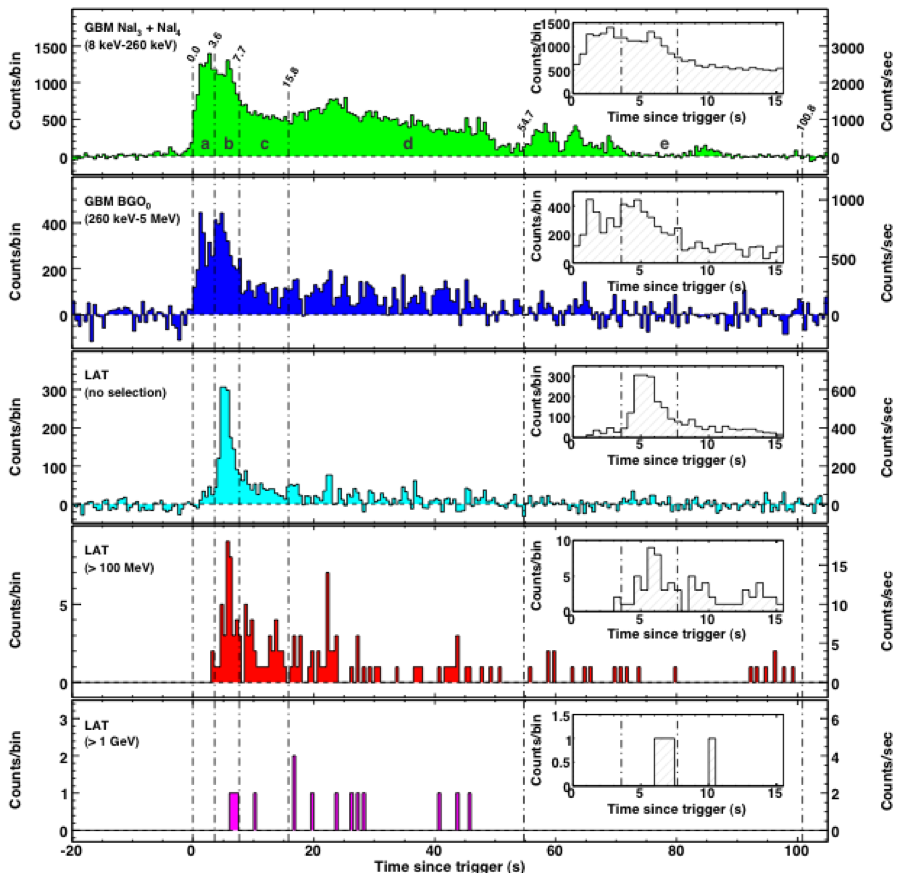}
\caption{
Light curves of GRB 080916C with the GBM (top three panels) and LAT
(bottom two panels).  The high energy LAT emission is delayed relative
to the lower energy GBM emission. From Abdo et
al. 2009 \cite{Abdo+080916C}.  }
\label{fig:LC080916C}
\end{figure}

LAT bursts have shown two common and interesting features: (1) delayed
emission compared to lower energy bands and (2) longer lasting prompt
emission compared to lower every bands.  The four brightest LAT bursts
are GRB 080916C \cite{Abdo+080916C}, GRB 090510 \cite{Abdo+090510,
Ackermann+090510}, GRB 090902B \cite{Abdo+090902B}, and GRB
090926A \cite{Ackermann+090926A}.  They have yielded hundreds of $>
100$ MeV photons each, and together with the lower energy GBM
observations, have given unprecedented broad-band spectra.  In GRB
080916C, the GeV emission appears only in a second pulse, delayed by
$\sim 4$ s relative to the first pulse (Figure \ref{fig:LC080916C}).
Such a delay is present also in short bursts, such as GRB 090510,
where it is a fraction of a second.  This soft-to-hard spectral
evolution is clearly seen in all four of these bright LAT bursts, and
to various degrees a similar behavior is seen in other weaker LAT
bursts.

In some bursts, such as GRB 080916C and several others, the broad-band
gamma-ray spectra consist of a simple Band-type broken power-law
function in all time bins.  In GRB 080916C the first pulse has a soft
high energy index disappearing at GeV energies, while the second and
subsequent pulses have harder high energy indices reaching into the
multi-GeV range.  In some other bursts, such as GRB
090510 \cite{Abdo+090510, Ackermann+090510} and GRB
090902B \cite{Abdo+090902B}, a second hard spectral component
extending above 10 GeV without any obvious break appears in addition
to common Band spectral component dominant in the lower 8 keV-10 MeV
band.

An exciting discovery, unanticipated by results from the Energetic
Gamma-Ray Experiment Telescope (EGRET) instrument on CGRO, was the
detection of high-energy emission from two short bursts (GRB 081024B
\cite{Abdo+081024B} and GRB 090510 \cite{Abdo+090510, Ackermann+090510}).  
Their general behavior (including a GeV delay) is qualitatively
similar to that of long bursts.  The ratio of detection rates of short
to long GRBs by LAT is $\sim 7\%$ (2 out of 27), which is
significantly smaller than the $\sim 20\%$ by GBM.

While the statistics on short GRBs are too small to draw firm
conclusions, so far, the ratio of the LAT fluence to the GBM fluence
is $\sim 100\%$ for the short bursts as compared to $\sim 5$--$60\%$
for the long bursts.  It is also noteworthy that, for both long and
short GRBs, the $\gtrsim 100$ MeV emission lasts longer than the GBM
emission in the $< 1$ MeV range.  The flux of the long-lived LAT
emission decays as a power law with time, which is more reminiscent of
the smooth temporal decay of the afterglow X-ray and optical fluxes
rather than the variable temporal structure in the prompt keV--MeV
flux.  This similarity in the smooth temporal evolution of the fluxes
in different wave bands has been detected most clearly in GRB 090510
\cite{DePasquale+10, Razzaque10} although this burst also requires a 
separate prompt component to the LAT emission \cite{He+11}.  This
short burst was at $z = 0.9$, and was jointly observed by the LAT,
GBM, BAT, XRT and UVOT.

The LAT detects only $\sim 10\%$ of the bursts triggered by the GBM
which were in the common GBM-LAT field of view.  This may be related
to the fact that the LAT-detected GRBs, both long and short, are
generally among the highest fluence bursts, as well as being among the
intrinsically most energetic GRBs.  For instance, GRB 080916C was at
$z = 4.35$ and had an isotropic-equivalent energy of $E_{\rm
iso} \approx 8.8\times 10^{54}$ ergs in gamma rays, the largest ever
measured from any burst \cite{Abdo+080916C}.  The long LAT bursts GRB
090902B \cite{Abdo+090902B} at $z = 1.82$ had $E_{\rm iso} \approx
3.6\times 10^{54}$ ergs, while GRB 090926A \cite{Ackermann+090926A} at
$z = 2.10$ had $E_{\rm iso} \approx 2.24\times 10^{54}$ ergs.  Even
the short burst GRB 090510 at $z = 0.903$ produced, within the first 2
s, an $E_{\rm iso} \approx 1.1\times 10^{53}$
ergs \cite{Ackermann+090510}.

\subsection{High redshift GRBs}
\label{subsec:highz}

Swift is detecting GRBs at higher redshift than previous missions due
to its higher sensitivity and rapid afterglow observations.  The
average redshift for the Swift GRBs is 2.3 compared to 1.2 for
previous observations.  Although statistics are poor, the highest
redshift GRBs are seen to have high luminosity, resulting in fluxes
well above the detection threshold.  Such bursts are also strong at
other wavelengths.  Table 1 presents optical data for the highest
redshift GRBs observed to date.  The optical brightness is given in
magnitudes in $R$, $I,$ $J$ and $K$ wavelength bands.  The bursts are
3--4 orders of magnitude brighter than typical galaxies at these
redshift.

\begin{figure}
\includegraphics[width=3.2in]{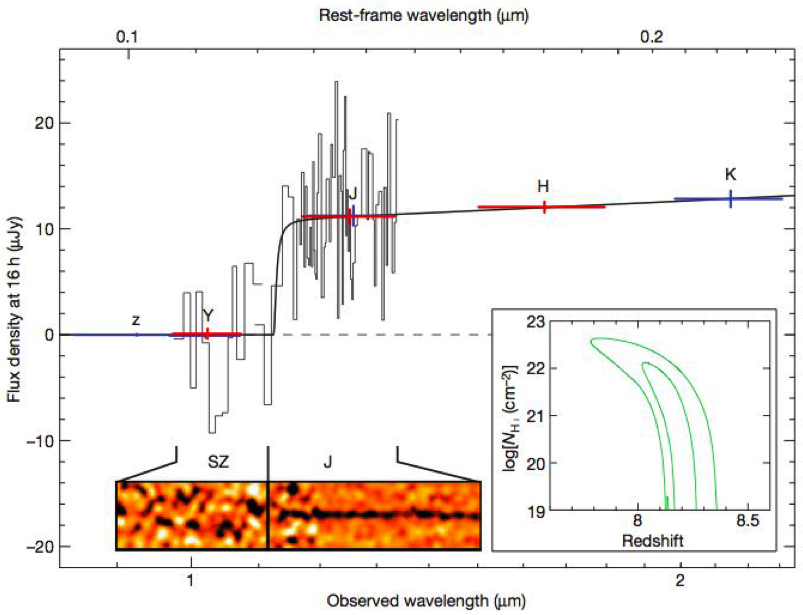}
\caption{
The composite spectrum of the GRB 090423 afterglow obtained with the
VLT. Also plotted are the sky-subtracted photometric data points
obtained using Gemini North \& South and VLT.  A model spectrum
showing the H I damping wing for a host galaxy with a hydrogen column
density of NH I $5\times 10^{21}$ cm$^{-2}$ at a redshift of $z =
8.23$ is also plotted (solid black line).  Inset, allowing for a wider
range in possible host NH I values gives the $1\sigma$ (68\%) and
$2\sigma$ confidence contours shown. From Tanvir et
al. 2009 \cite{Tanvir+09}.
}
\label{fig:090423}
\end{figure}

To date, Swift has seen seven GRBs with $z > 5$.  The highest $z$ GRBs
so far are $z = 6.3$ (GRB 050904 \cite{Kawai+06}), $z = 6.7$ (GRB
080913 \cite{Greiner+09}), $z = 8.2$ (GRB 090423 \cite{Tanvir+09,
Salvaterra+09}), and $z = 9.4$ (GRB 090429B \cite{Cucchiara+11}).
Figure \ref{fig:090423} shows a spectrum of the current record holder
with a spectroscopic redshift, GRB 090423 \cite{Tanvir+09}.  Because
GRBs are so bright across the electromagnetic spectrum, they are
excellent tools for studying the high redshift universe. The
information that can be provided by GRBs is complementary to what can
be learned by studies of galaxies and quasars.  For example, at $z >
8$ it is expected that the bulk of the star formation activity is
taking place in small galaxies below the detection limits of HST.

\begin{figure}
\includegraphics[width=3.in]{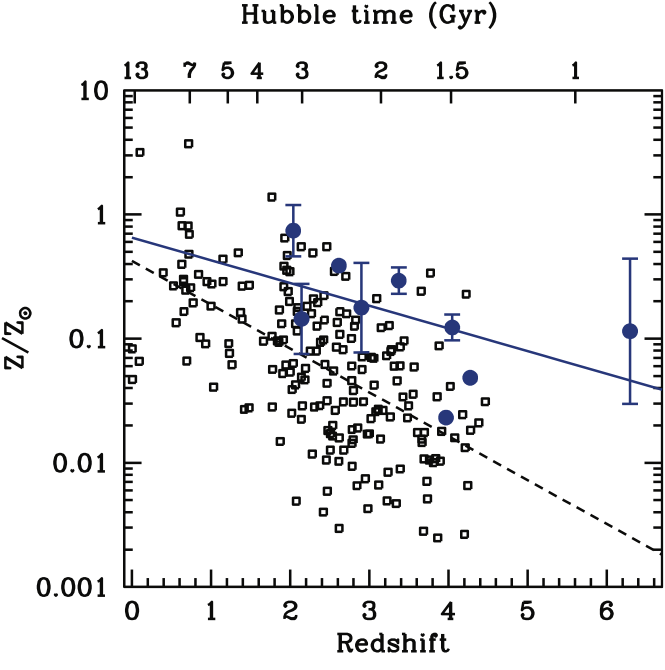}
\caption{
Redshift evolution of the metallicity (represented here by the ratio
of oxygen to hydrogen abundance) relative to solar values, for GRBs
shown with blue dots and active galaxy quasars show with open circles.
The abundances are determined from spectral absorption lines in the
continuum radiation.  GRB lines are predominantly from the gas in the
host galaxy in the star-forming region near the explosion, whereas
quasars are of random lines-of-sight through the galaxy.  The GRB
metallicity is on average $\sim 5$ times larger than in QSO. These are
based on damped Lyman alpha (DLA) spectral features.  The upper
horizontal x-axis indicates the age of the Universe (Hubble
time). From Savaglio et al. 2006 \cite{Savaglio+06}.}
\label{fig:metallicity}
\end{figure}

The chemical evolution of the universe can be studied with GRBs as
illustrated in Figure \ref{fig:metallicity} \cite{Savaglio+06}.  GRBs
provide data to higher redshift than active galaxies.  The metallicity
bias of GRBs is currently contradictory.  Plots like
Figure \ref{fig:metallicity} show GRBs in higher metallicity star
forming regions of galaxies.  However, comparing star forming regions,
GRBs actually tend to favor ones with lower
metallicity \cite{Levesques+10}.

\begin{figure}
\includegraphics[width=3.2in]{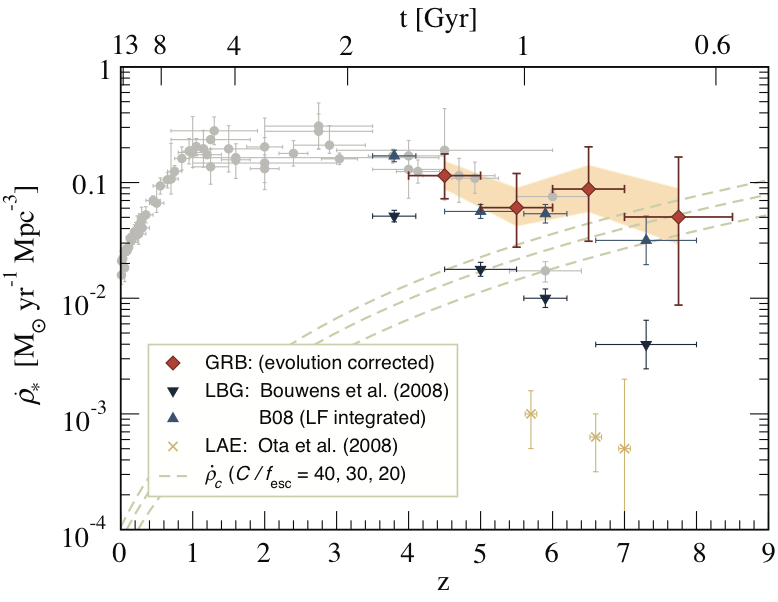}
\caption{
Cosmic star formation history. Shown are the data compiled in Hopkins
\& Beacom \cite{Hopkins+06} (light circles) and contributions from Lyα emitters
(LAE) \cite{Ota+08}. Recent LBG data are shown for two UV LF
integrations: down to $0.2 L_*$ (down triangles; as given in Bouwens
et al. \cite{Bouwens+08}) and complete $z=3$ (up triangles).  Swift
GRB-inferred rates are diamonds, with the shaded band showing the
range of values resulting from varying the evolutionary parameters.
Also shown is the critical density derivative from Madau et
al. \cite{Madau+99} for $C/f_{esc} = 40,$ 30, 20 (dashed lines, top to
bottom).  From Kistler et al. 2009 \cite{Kistler+09}. See also study
by Robertson \& Ellis \cite{Robertson+12}.}
\label{fig:starformation}
\end{figure}

Another way that GRBs are contributing to our understanding of the
high-redshift universe is in the determination of the star formation
history (Figure \ref{fig:starformation}).  LGRBs are the endpoints of
the lives of massive stars and their rate is therefore approximately
proportional to the star formation rate.  They give information at
high redshift where the rate is highly uncertain.  There may be
evolutionary biases, such as a dependence of LGRBs on the metallicity
of host galaxies, so studies relating them to the star formation rate
must include these factors into account \cite{Kistler+09,
Robertson+12}.

\begin{table}[h]
\caption{Optical Observations of High Redshift GRBs}
\begin{tabular}{llll}
\hline
$z$ & $t_{\rm LB}$ (Gyr) & GRB & Brightness \\
\hline
9.4 & 13.1 & 090429B &  K = 19 @ 3hr \\
8.2 & 13.0 & 090423 &	K = 20 @ 20 min \\
6.7 & 12.8 & 080913 &	K = 19 @ 10 min \\ 
6.3 & 12.8 & 050904 &	J = 18 @ 3 hr \\
5.6 & 12.6 & 060927 &	I = 16 @ 2 min \\
5.3 & 12.6 & 050814 &	K = 18 @ 23 hr \\
5.11 & 12.5 & 060522 &	K = 21 @ 1.5 hr \\
\hline
\end{tabular}
\end{table}

\subsection{GRB-Supernova connection}
\label{subsec:grbsn}

On 18 February 2006 Swift detected the remarkable burst GRB 060218
that provided considerable new information on the connection between
SNe and GRBs.  It was longer (35 min) and softer than any previous
burst, and was associated with SN 2006aj at only $z=0.033$. SN 2006aj
was a (core-collapse) SN Ib/c with an isotropic energy equivalent of a
few $10^{49}$ erg, thus underluminous compared to the overall energy
distribution for long GRBs. The spectral peak in prompt emission at
$\sim 5$ keV places GRB 060218 in the X-ray flash category of GRBs
\cite{Robertson+12}, the first such association for a GRB-SN event.  
Combined BAT-XRT-UVOT observations provided the first direct
observation of shock-breakout in a SN \cite{Campana+06}.  This is
inferred from the evolution of a soft thermal component in the X- ray
and UV spectra, and early-time luminosity variations. Concerning the
SN, SN 2006aj was dimmer by a factor $\sim 2$ than the previous SNe
associated with GRBs, but still $\sim 2$--3 times brighter than normal
SN Ic not associated with GRBs \cite{Pian+06, Mazzali+06}. GRB 060218
was an underluminous burst, as were two of the other three previous
cases.  Because of the low luminosity, these events are only detected
when nearby and are therefore rare. However, they are actually $\sim
5$--10 times more common in the universe than normal
GRBs \cite{Soderberg+08} {and form a distinct component in the
luminosity function \cite{Liang+07}.}

GRB 060218 joins 4 other previous cases of GRBs from other missions
that were associated with specific supernovae (980425, 021211, 030329,
031203).  In addition, there are several other Swift cases (050525A,
091127, 100316D, 101219B and 120422A).  Table 2 gives a summary of all
long GRBs that have had either specific supernovae or features in
their afterglow lightcurves suggestive of the presence of a supernova.
Many of the nearby GRBs with specific supernovae are in the
underluminous class like GRB 060218.  There are also a few cases
listed in the table of nearby GRBs that had supernova searches,
without a detection.  Two of these, GRB 060505 and 060614 are
discussed further in section \ref{subsec:oddballs}.

\begin{table}[htb]
\caption{Nearby GRBs and Supernova Detections or Limits}
\begin{tabular}{llll}
\hline
GRB	& Redshift, $z$	& Type\footnote{UL = underluminous, XRF = X-Ray Flash}	& SN Search \\
\hline
980425	& 0.0085 & long-UL	& SN 1998bw \\
020903	& 0.251	& XRF-UL	& LC bump \& spectrum \\
021211	& 1.006	& long-UL	& SN 2002lt \\
031203	& 0.105	& long-UL	& SN 2003lw \\
030329	& 0.168	& long	        & SN 2003dh \\
050525A	& 0.606	& long-~UL	& SN 2005nc \\
060218	& 0.033	& XRF-UL	& SN 2006aj \\
091127	& 0.49	& long-UL	& SN 2009nz \\
100316D	& 0.059	& long-UL	& SN 2010bh \\
101219B	& 0.55	& long-UL	& SN 2010ma \\
120422A	& 0.283	& long-UL	& NS 2012bz \\
011121	& 0.36	& long-UL	& LC bump \& spectrum \\
050826	& 0.297	& long-UL	& LC bump \\
060729	& 0.54	& long-~UL	& LC bump \\
090618	& 0.54	& long	        & LC bump \\
080120	&	& long	        & LC bump GROND \\ 
081007	&	& long	        & LC bump GROND \\
090424	&	& long	        & LC bump GROND \\
100902A	&	& long	        & LC bump GROND \\
110402A	&	& long	        & LC bump GROND \\
        &       &               &		\\	
040701	& 0.215	& XRF-UL	& no SN ($<0.1$ SN98bw) \\
060505	& 0.089	& ``long''-UL	& no SN ($<0.004$ SN98bw) \\
060614	& 0.125	& ``long''	& no SN ($<0.01$ SN 98bw) \\
101225A	& 0.40	& long	        & no SN (GCN 11522) \\
\hline
\end{tabular}
\end{table}

\subsection{Short GRBs}
\label{subsec:sgrb}

At the time of Swift’s launch, the greatest mystery of GRB astronomy
was the nature of short-duration, hard-spectrum bursts (SGRBs).
Although more than 50 long GRBs had afterglow detections, no afterglow
had been found for any short burst.  In summer 2005 Swift and HETE-2,
precisely located three short bursts for which afterglow observations
were obtained leading to a breakthrough in our understanding of short
bursts \cite{Gehrels+05, Bloom+06, Barthelmy+05b, Berger+05, Fox+05,
Hjorth+05, Villasenor+05}. As of 2012, BAT has detected 65 SGRBs,
$80\%$ of which have XRT detections, and 15 of which have redshifts.

In contrast to long bursts, the evidence is that SGRBs typically
originate in host galaxies with a wide range of star formation
properties, including low formation rate.  Their host properties are
substantially different than those of long bursts \cite{Leibler+10,
Fong+10} indicating a different origin.  Also, nearby SGRBs show no
evidence for simultaneous supernovae (\cite{Nakar07}; and references
therein), very different than long bursts.  Taken together, these
results support the interpretation that SGRBs arise from an old
populations of stars and are due to mergers of compact binaries (i.e.,
double neutron star or neutron star - black hole) \cite{Nakar07,
Eichler+89, Paczynski91}.  {It is possible that there is a
subpopulation of high-redshift, high-luminosity short GRBs with a
different origin \cite{Zhang+09, Bromberg+12}.}

Measurements or constraining limits on beaming from light curve break
searches have been hard to come by with the typically weak afterglow
of SGRBs.  With large uncertainties associated with small number
statistics, the distribution of beaming angles for SGRBs appears to
range from $\sim 5^\circ$ to $> 25^\circ$ \cite{Burrows+06, Fong+12},
roughly consistent but perhaps somewhat larger than that of LGRBs.
Swift observations also revealed long ($\sim 100$ s) ``tails'' with
softer spectra than the first episode following the prompt emission
for about $25\%$ of short bursts \cite{Norris+06, Gehrels+06}.  Swift
localization of a short GRB helps narrow the search window for
gravitational waves from that GRB \cite{Finn+99}.  Detection of
gravitational waves from a Swift GRB would lead to great scientific
payoff for merger physics, progenitor types, and NS equations of
state.

\subsection{Unusual GRBs and transients}
\label{subsec:oddballs}

Here is a list of unusual transients detected by
Swift \cite{Gehrels+12b}.

{\em Short GRB 050925:} This unusual burst triggered the BAT with a
single peaked outburst of duration $T90 = 70$ ms \cite{Holland+05,
Markwardt+05}.  It occurred near the galactic plane, and nothing was
seen in the UVOT.  However, the $V$-band extinction toward the source
was $AV =7.05$ mag.  The XRT spectra and lightcurve show no
significant X-ray emission in the field, suggesting that any X-ray
counterpart to this burst was faint. Markwardt et
al. (2005) \cite{Markwardt+05} were able to fit a power law spectrum
to the BAT data, with a photon index $1.74\pm 0.17$; they found a
better fit could be obtained with a blackbody spectrum $kT = 15.4 \pm
1.5$ keV, a value consistent with small-flare events from SGRs. The
low galactic latitude and soft spectrum indicate a possible galactic
source or SGR.

\begin{figure}
\includegraphics[width=3.2in]{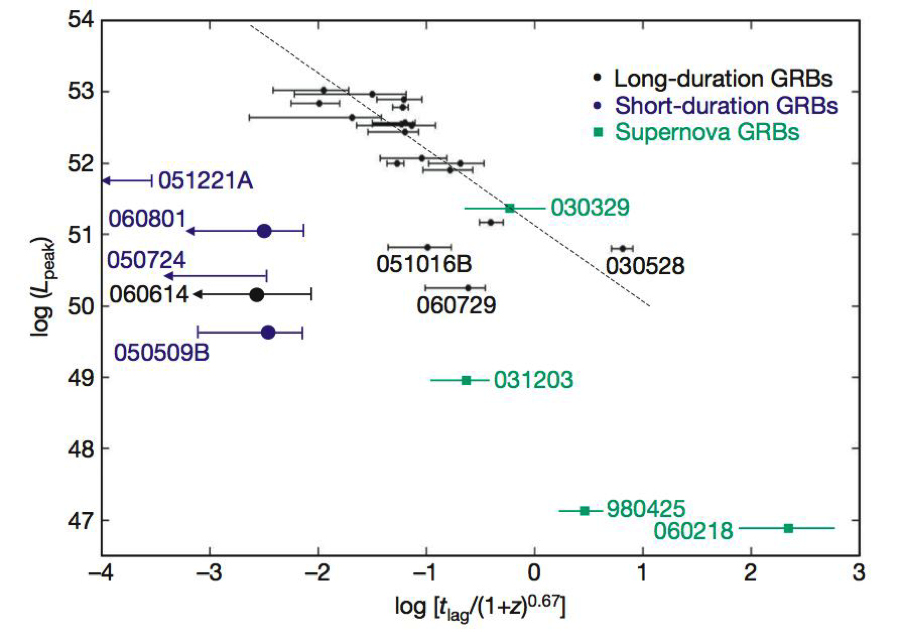}
\caption{
Spectral lag as a function of peak luminosity showing GRB 060614 in
the region of short-duration GRBs \cite{Gehrels+06}. }
\label{fig:lag}
\end{figure}

{\em GRBs 060505 \& 060614:} GRB 060505 and GRB 060614 were nearby
GRBs ($z = 0.089$ and 0.125, resp.) with no coincident SN (Table 2),
to deep limits \cite{Fynbo+06}.  GRB 060614 was bright (15--150 keV
fluence of $2.2\times 10^{-5}$ erg cm$^{-2}$), and, with a $T90$ of
102 s, seemed to be a secure long GRB.  Host galaxies were
found \cite{Fynbo+06, Gal-Yam+06, dellaValle+06} and deep searches
were made for coincident SNe.  All other well-observed nearby GRBs
have had SNe, but GRB 060614 did not to limits $> 100$ times fainter
than previous detections \cite{Markwardt+05, Fynbo+06, Gal-Yam+06}.
GRB 060614 shares some characteristics with SGRBs \cite{Gehrels+06}.
The BAT light curve shows an initial short hard flare lasting $\sim 5$
s, followed by an extended softer episode, $\sim 100$ s. The light
curve is similar to some Swift SGRBs and a subclass of BATSE
SGRBs \cite{Norris+06}.  GRB 060614 also falls in the same region of
the lag-luminosity diagram as SGRBs (Figure \ref{fig:lag}). Thus GRB
060614 is problematic to classify.  It is a LGRB by the traditional
definition, but lacks an associated SN. It shares some similarities
with SGRBs, but the soft episode is brighter, which would be difficult
to account for in the NS-NS merger scenario.

{\em Hostless GRB 070125:} There is no obvious host galaxy for GRB
070125.  Deep ground-based imaging reveals no host to $R > 25.4$ mag.
Cenko et al. (2008) \cite{Cenko+10} present an analysis of
spectroscopic data which reveals weak Mg II lines indicative of halo
gas.  In the field are two blue galaxies offset by $\gtrsim 27$ kpc at
$z = 1.55$.  If there is an association with one of them, it would
imply a velocity $\sim 10^4$ km s$^{-1}$ over a $\sim 20$ Myr lifetime
of the massive progenitor.  The only known way of achieving this would
have been a prior close interaction with a massive BH.  However, this
interpretation was muddied by Chandra et al. (2008) \cite{Chandra+08},
who inferred a dense environment, based on bright, self-absorbed radio
afterglow.  They proposed a scenario in which the high density
material lies close to the explosion site, and the lower density
material further away.  They note GRB 070125 was one of the brightest
GRBs ever detected, with an isotropic release of $10^{54}$ erg (by
comparison, $M_\odot c^2 \approx 2 \times 10^{54}$ erg).  The prompt
emission from GRB 070125 was also seen by Suzaku/WAM \cite{Onda+10}.

{\em Galactic GRB 070610:} Discovered initially as GRB 070610, this
object, now dubbed Swift J195509.6+261406 (Sw1955+26), is thought to
represent a member of a relatively new class of fast X-ray nova
containing a BH.  It had a duration of $\sim 5$ s, and also shows
large variability.  Kasliwal et al.\ (2008) \cite{Kasliwal+08} discuss
several possibilities for this source, and propose an analogy with
V4641 Sgr, an unusual BH binary which had a major outburst in
1999 \cite{Zand+00}.  V4641 Sgr is a binary with a B9 III star
orbiting a $\sim 9 M_\odot$ BH \cite{Orosz+01} and also exhibited
strong and fast X-ray and optical variability.  The analog is
imperfect in that the normal star in Sw1955+26 is a cool dwarf rather
than a B9 giant, suggesting a physical origin for the bursting
behavior in the accretion disk and/or jet rather than the mass donor
star.  A $\sim 5$ s burst is certainly distinct from the $\sim$ month
long fast-rise exponential decay seen in systems like A0620- 00 in
1975 \cite{Tanaka+96, Remillard+06} which are thought to be due to a
large scale storage and dumping of material in an accretion
disk \cite{Cannizzo+95, Cannizzo98}, and may be more in line with
either a disk-corona \cite{Nayakshin+00} or disk-jet
instability \cite{McKinney+09}.  Rea et al. (2011) \cite{Rea+11}
derive stringent upper limits on the quiescent X-ray emission from
Sw1955+26 using a $\sim 63$ ks Chandra observation, and use this to
argue against a magnetar interpretation.  Simon et
al. (2012) \cite{Simon+12} present an analysis of optical emission
during the 2007 outburst and find the optical emission manifests as
spikes which decrease in duration as the burst progresses. They show
that the emission can be explained by pure synchrotron emission. The
main part of the optical outburst following the Swift/BAT trigger
lasted $\sim 0.3$ d, with subsequent echo outbursts between 1 and 3
day, as in X-ray novae.  The main part of the outburst has an
exponential decay reminiscent of the 1975 outburst in A0620-00, but
with a much shorter duration.  If the global accretion disk limit
cycle \cite{Cannizzo+95, Dubus+01} the mechanism for this outburst, it
suggests a much shorter orbital period than the 7.75 hr for
A0620-00 \cite{McClintock+86}.

\begin{figure}
\includegraphics[width=3.2in]{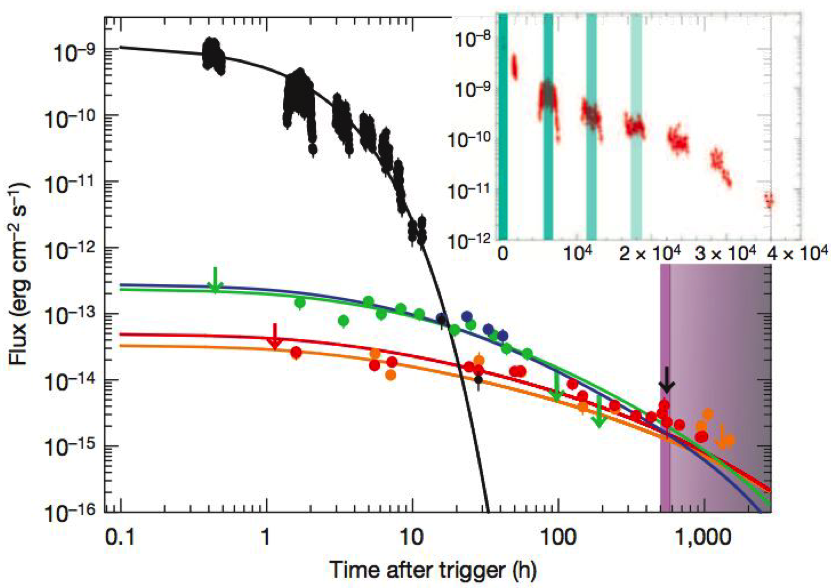}
\caption{
Light curves of GRB 101225 in five energy bands: X-rays at 1 keV
(black), UV at 2030 angstroms (green) and 2634 angstroms (blue), and
optical at 6400 angstroms ($R$ band, red) and 7700 angstroms ($I$
band, orange) \cite{Campana+11}.  The inset (with t in seconds) shows
the Swift/XRT light curve.  Shaded regions highlight the periastron
passages calculated using a tidal disruption
model \cite{Campana+11}. }
\label{fig:101225}
\end{figure}

{\em GRB 101225 Christmas burst:} GRB 101225 was quite unusual: it had
a $T90 > 1700$ s and exhibited a curving decay when plotted in the
traditional $log F$-$log t$ coordinates. The total BAT fluence was
$\gtrsim 3 \times 10^{-6}$ erg cm$^{-2}$. The XRT and UVOT found a
bright, long-lasting counterpart.  Ground based telescopes followed
the event, mainly in $R$ and $I$, and failed to detect any spectral
features.  At later times a color change from blue to red was seen;
HST observations at 20 d found a very red object with no apparent
host. Observations from the Spanish Gran Telescopio Canarias at 180 d
detected GRB 101225 at $gAB = 27.21 \pm 0.27$ mag and $rAB = 26.90 \pm
0.14$ \cite{Thone+11}.  Considered together, these characteristics are
unique to this burst (Figure \ref{fig:101225}), and led Campana et
al. (2011) \cite{Campana+11} to propose that it was caused by a minor
body like an asteroid or comet becoming disrupted and accreted by a
NS.  Depending on its composition, the tidal disruption radius would
be $\sim 10^5$--$10^6$ km. Campana et al. find an adequate fit to the
light curve by positing a $\sim 5 \times 10^{20}$ g asteroid with a
periastron radius $\sim 9000$ km.  If half the asteroid mass is
accreted they derive a total fluence $4 \times 10^{-5}$ erg cm$^{-2}$
and distance $\sim 3$ kpc.  Th\"one et al. (2011) \cite{Thone+11}
offer a different explanation for GRB 101225 --- the merger of a He
star and a NS leading to a concomitant SN.  They derive a
pseudo-redshift $z=0.33$ by fitting the spectral-energy distribution
and light curve of the optical emission with a GRB-SN template.  Thus
in their interpretation the event was much more distant and energetic.
They argue for the presence of a faint, unresolved galaxy in deep
optical observations, and fit the long-term light curve with a
template of the broad-line type Ic SN 1998bw associated with GRB
980425.  If their distance is correct as well as their interpretation
of a component emerging at 10 d as being a SN, then its absolute peak
magnitude $M V_{abs} = -16.7$ mag would make it the faintest SN
associated with a long GRB.  The isotropic-equivalent energy release
at $z = 0.33$ would be $> 1.4 \times 10^{51}$ erg which is typical of
other long GRBs but greater than most other low-redshift GRBs
associated with SNe.

\section{Theory and Radiation Modeling}
\label{sec:theory}

Gamma-ray bursts are thought to be the endpoints of evolution of
massive, $\gtrsim 30 M_{\odot}$, stars (favorite for long
GRBs) \cite{Gehrels+09} or neutron star-neutron star/black hole close
binaries (favorite for short GRBs) \cite{Narayan+92, Nakar07,
Eichler+89, Paczynski91}.  In either case a black hole and a
short-lived accretion disc are thought to be formed after the core
collapse (long GRBs) or binary merger (short GRBs).  In some cases a
highly magnetized neutron star, called a magnetar, may form as an
intermediate state which eventually collapses to a black
hole {\cite{Vietri+99}.}  Gravitational binding energy of the
system is released within a short time scale which eventually produces
the GRB phenomenon when some fraction of the gravitational energy
converts to gamma rays.  This is done via accretion of material/gas
onto the black hole.

The compact core of a few solar mass with the Schwarzschild radius
$r_g \sim 10^6(M_{BH}/3 M_\odot)$ cm forms the central engine of a
GRB.  Infalling gas onto the newly-formed black hole is accreted with
high efficiency.  Most of the gravitational energy, $\sim 10^{54}$
ergs, is radiated in $\sim 10$ MeV thermal neutrinos as in
core-collapse SNe and a lesser amount in gravitational
wave. {Only a small fraction, $\sim 10^{51}$ erg, is converted to
a GRB fireball,} which eventually produces a highly-relativistic
bi-polar jet and is responsible for the observed $\gamma$
rays \cite{Meszaros+12, Gehrels+12}.

As discussed previously, observations of supernovae associated with
long GRBs provide strong support for the core-collapse schenario for
this class of GRBs.  The binary merger scenario of short GRBs has not
been confirmed, but remains the most popular model.

\subsection{GRB fireball model}
\label{subsec:fireball}

The GRB fireball is produced when large amount of energy is released
within a short time scale.  The isotropic-equivalent energy is
$E_{iso} = (4\pi/\Omega_j)E_j$, where $\Omega_j$ is the jet solid
angle.  The average solid angle $\langle \Omega_j \rangle/4\pi \sim
1/500$ or {the jet half-opening angle
$\langle \theta_j \rangle \sim 1/10$} \cite{Frail+01, Cenko+09,
Gehrels+09}.  The luminosity of the outflowing material is
$L_{iso} \sim E_{iso}/t$, where $t$ is the typical duration of a long
GRB.  The temperature of the outflowing material at the base of the
jet, with an inner radius of an accretion disc $r_0\sim 10^7$ cm, is
$kT = (L_{iso}/4\pi r_0^2 ca) \sim$ few MeV.  Here $a = 7.6\times
10^{-15}$ ergs cm$^{-3}$ K$^{-4}$ is the radiation density constant.
The fireball contains gamma rays, electron-positron pairs and
relatively fewer baryons (mostly protons).  Due to the radiation
pressure, the optically thick hot plasma expands and overcome the
gravitational pressure and forms the relativistic
jet \cite{Meszaros+93}.

An initially thermal energy dominated GRB jet becomes kinetic energy
dominated when the baryons carry the bulk of the initial fireball
energy at a saturation radius $>r_0$ \cite{Shemi+90}.  The bulk
Lorentz factor of the jet initially increases as $\Gamma (r) \sim
r/r_0$ and becomes constant at a value $\Gamma \sim \eta \sim
10^2$--$10^3$, where $\eta \sim L_{iso}/{\dot M}c^2$ is defined as the
ratio beteen the total energy to mass flow and is called the baryon
loading of the fireball.  The radius at which the fireball attains
this coasting Lorentz factor is $\sim r_0\eta$ \cite{Meszaros+00,
Rossi+06}.

\subsection{Prompt keV--MeV emission}
\label{subsec:prompt}

The prompt $\gamma$-rays from a GRB are thought to be emitted when a
significant fraction of the fireball kinetic energy is converted back
to radiation energy.  The internal shocks model is currently the most
popular model for prompt $\gamma$-ray emission \cite{Rees+94}.
Collisions between shells of plasma or fireball produce shocks and a
fraction of the bulk kinetic energy is converted into random particle
energy which is then radiated, via synchrotron mechanism e.g., as the
observed non-thermal $\gamma$ rays.  Given a timescale $t_v$ from
observations of highly variable prompt $\gamma$-ray emission, internal
shock collisions take place, due to a difference in Lorentz factor
$\Delta \Gamma \sim \Gamma$ between the shells, at a radius
$r_i \sim \Gamma^2 ct_v \sim 3\times 10^{13} (\Gamma/316)^2
(t_v/0.01~{\rm s})$ cm. This is typically much larger than the
photospheric radius of a {baryon-dominated GRB jet,} given by
$r_{ph} \sim \sigma_T L_{iso}/(4\pi \Gamma^3 m_pc^3) \sim 4\times
10^{12} (L_{iso}/10^{52}~{\rm erg~s}^{-1}) (\Gamma/316)^3$
cm {(see, however, \cite{Veres+12} for different scaling of the
photosphere)}.  The observed pulse structure can be produced in
colliding shells model \cite{Kobayashi+97, Dermer04}. Note that the
variability time scale can be as small as $\sim 1$ ms, corresponding
to the size scale of the jet base $r_0$.

The internal shocks are semi-relativistic, with Lorentz factor
$\Gamma_{sh} \sim$ few.  A fraction of the shock energy is thought to
amplify magnetic field, via some kind of instabilities, up to an
energy that is at some equipartition level $\epsilon_B$ with the jet
luminosity $L_{iso}$ \cite{Medvedev+99, Inoue+11}. Another fraction of
the shock energy is thought to accelerate electrons, via a Fermi
mechanism, to relativistic energies.  In a fast-cooling scenario,
where electrons lose most of their energy to the observed radiation,
e.g.\ via synchrotron/Compton mechanism, within the dynamic time
scale, the prompt $\gamma$-ray luminosity $L_{\gamma,iso}$ can be used
to infer $L_{iso}$ (and eventually $E_{iso}$ from the duration of the
burst) as $L_{iso} = L_{\gamma,iso}/\epsilon_e$.  The magnetic field
in the jet frame can also be inferred from the prompt $\gamma$-ray
observations as $B^\prime = (2\epsilon_B L_{\gamma,iso}/\epsilon_e
r_i^2 \Gamma^2 c)^{1/2} \sim 2\times 10^4
(\epsilon_B/\epsilon_e)^{1/2} (L_{\gamma,iso}/10^{51}~{\rm
erg~s}^{-1})^{1/2} (\Gamma/316)^{-3} (t_v/0.01~{\rm s})^{-1}$
Gauss \cite{Razzaque+04}.  With a minimum Lorentz factor of the
electrons, $\gamma_e \sim m_p/m_e$, characteristic synchrotron
radiation peak can explain the observed $\varepsilon_{pk} \sim $ few
hundred keV typical $\nu F_\nu$ peak of the $\gamma$-ray spectrum.
The power-law part of the observed Band spectra above
$\varepsilon_{pk}$ can be explained as due to synchrotron radiation
from the power-law distributed electron Lorentz factor,
$n(\gamma_e) \sim \gamma_e^{-p}$, as
$n(\varepsilon) \sim \varepsilon^{-p/2 - 1}$.

The synchrotron shock model as described above has difficulties to
explain the observed $\gamma$-ray spectrum below $\varepsilon_{pk}$ in
a large fraction of the bursts \cite{Preece+98, Kaneko+06,
Goldstein+12}.  The fast-cooling synchrotron model is expected to
produce a spectrum $n(\varepsilon) \sim \varepsilon^{-3/2}$ below
$\varepsilon_{pk}$ as contrary to $\varepsilon^{-1}$, typical observed
value.  A harder, $\varepsilon^{-2/3}$, synchrotron spectrum from
electrons in random megnetic field is often called the ``synchrotron
death line'' \cite{Preece+98} which too has been observed violated in
a sizeable fraction of GRBs.

There have been many attempts to solve this Band spectrum crisis
within the synchro-Compton mechanism.  Explanations include
synchrotron self-absorption in the keV range \cite{Granot+00}, Compton
scattering of optical photons to keV energies \cite{Panaitescu+00},
low-pitch angle scattering or jitter radiation \cite{Medvedev00,
Medvedev06}, time-dependent acceleration \cite{Lloyd-Ronning+02} as
well as mixing of keV photons with axion-like
particles \cite{Mena+11}.  Photospheric thermal emission, as we
decribe next, holds promise to explain the hard Band spectrum below
$\varepsilon_{pk}$.

\subsection{Photospheric emission}
\label{subsec:photo}

In case of a baryon-loaded GRB fireball, as we discussed before, most
of the thermal energy converts to kinetic energy at a radius $r >
r_0\eta$. The photospheric radius $r_{ph}$ is also typically much
larger than $r_0\eta$.  {In case the fireball is highly baryon
dominated $r_{ph}$ is smaller and could become comparable to
$r_0\eta$. A significant fraction of the thermal energy can be
released from the fireball in this case \cite{Meszaros+00}.}  Moreover
dissipation of kinetic energy at or below the photosphere, by shocks
or other mechanisms such as neutron-proton decoupling and
interactions \cite{Derishev99, Rossi+06, Razzaque+06, Beloborodov10},
could also enhance thermal radiation for moderately baryon-loaded
fireballs as well as explain highly-efficient $\gamma$-ray emission
from GRBs, which is difficult to explain with internal shocks.

Simple photospheric models should produce thermal emission peaking at
$\sim 1$ MeV, the temperature of the fireball at the base of the jet,
which is in the range of observed Band peaks \cite{Rees+05, Pe'er+06}.
This leads to a plausible physical explanations for the
Amati \cite{Amati+02} or Ghirlanda \cite{Ghirlanda+04} relations
between the $\nu F_\nu$ peak energy and the total $\gamma$-ray
energy \cite{Rees+05, Thompson+06}.  {The temperature of the
photosphere can be much below $\sim 1$ MeV in more complicated models,
see e.g.\ \cite{Veres+12}.} Neutron-proton decoupling, their
interactions and resulting radiation from cascade particles at the
photosphere or Compton scattering of thermal photons by relativistic
electrons in sub-photospheric shocks have also been explored to
produce the phenomenological Band spectrum.  These processes explain
the power-law spectrum above $\varepsilon_{pk}$.  The ``synchrotron
death line'' issue \cite{Preece+98} below $\varepsilon_{pk}$ can be
avoided in photospheric models.

Recent modeling of Fermi data from GRB 090902B provides some evidence
of photospheric emission being dominant in this burst \cite{Pe'er+12}.
In some other Fermi bursts a photospheric component in addition to the
Band or Comptonized spectrum seems to improve fits, although thermal
emission is not dominant in these bursts \cite{Guiriec+10,
Guiriec+11}.

\subsection{Poynting flux dominated GRB model}
\label{subsec:magnetic}

Fermi observations of GRBs over a broad energy range and non-detection
of a strong photospheric component in almost all bursts, {and
specially after GRB 080916C \cite{Abdo+080916C, Zhang+09b},} have
renewed interests in magnetic or Poynting-flux dominated GRB jets, .
A large dissipation radius, $r \sim 10^{15}$ cm, yet high efficiency
inferred from modeling Fermi data can also be explained more naturally
with magnetic models.  Particles are accelerated in the reconnection
region of the magnetic field lines and produce the observed radiation.
Magnetic models can be generalized into two categories. In the first
category the baryons are completely absent in the jet or dynamically
negligible, at least initially {\cite{Usov94, Meszaros+97b,
Lyutikov+03},} and in the second category the baryon load is
significant but dynamically sub-dominant relative to the magnetic
stresses {\cite{Thompson94, Drenkhahn02, Drenkhahn+02,
Thompson06}.} A hybrid model, dubbed ICMART \cite{Zhang+11} involves
an initially Poynting-flux dominated outflow with few baryons that
lead to internal shocks when the magnetic energy subsides at a large
radius.

\subsection{GeV emission and plausible interpretations}
\label{subsec:gev}

\begin{figure}
\includegraphics[width=3.2in]{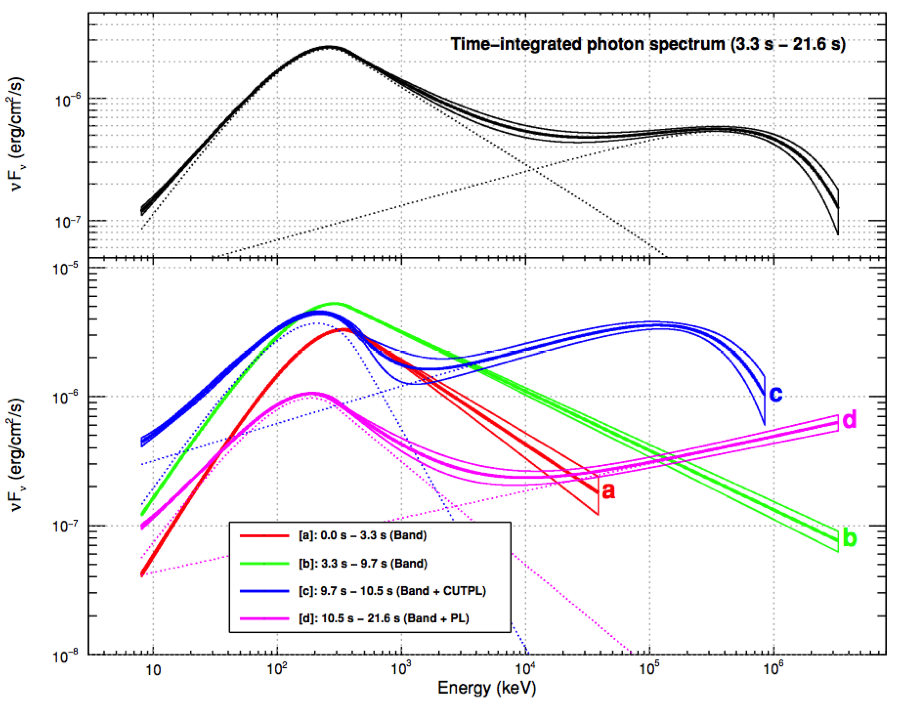}
\caption{
Prompt emission spectra of GRB 090926A at different time intervals
(bottom panel) as well as during the entire prompt phase (top panel).
A power-law component in addition to the Band spectrum, as also found
in other bright LAT-detected GRBs, is required to fit spectra.  A
cutoff in the PL component at few GeV has been detected in this GRB
and in GRB 110731A \cite{Ackermann+110731A}. From Abdo et
al. \cite{Ackermann+090926A}. }
\label{fig:spectra090926A}
\end{figure}

The origin of $\gtrsim 100$~MeV radiation from GRBs and associated
additional spectral component (Figure \ref{fig:spectra090926A}) has
been intensely debated in recent years.  In the context of the internal
shock scenario, Synchrotron-Self-Compton (SSC) scattering of soft
target photons by relativistic electrons is the first choice to
produce the hard power-law component in a simple one-zone
model \cite{Wang+09, Bosnjak+09, Toma+09, Ackermann+090510, Toma+11}.
The observed time delay (Figure \ref{fig:LC080916C}) between the
keV--MeV radiation and $\gtrsim 100$~MeV radiation, however, is longer
than the expected delay for the SSC emission, which can be of the
order of the variability time scale ($\lesssim 1$~s for long bursts)
at best.  Compton radiation from late internal shocks, with much
longer variability time at GeV than in keV--MeV, can explain delayed
LAT onset in some cases~\cite{Bosnjak+09, Toma+11}.

\begin{figure}
\includegraphics[width=3.in]{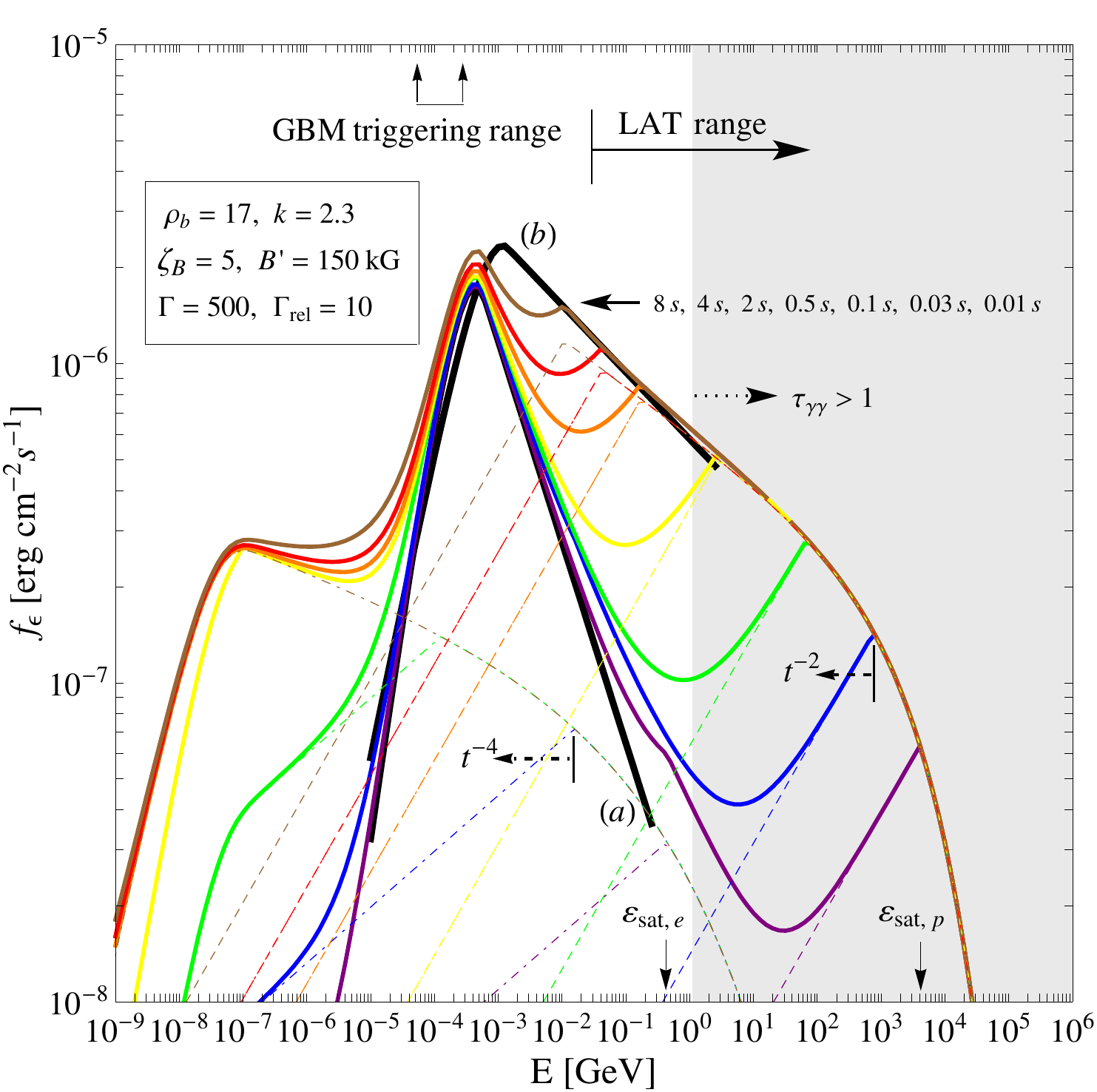}
\caption{ 
Proton-synchrotron radiation model for delayed onset of LAT emission
in GRB 080916C.  The delay is caused by a build-up of
proton-synchrotron flux and a shift of cooling frequency from very
high eenrgies to the LAT range, with time.  GBM radiationn is assumed
to be dominated by electron-synchrotron radiation.
From~\cite{Razzaque+10}. }
\label{fig:psynchrotron}
\end{figure}

\begin{figure}
\includegraphics[width=3.in]{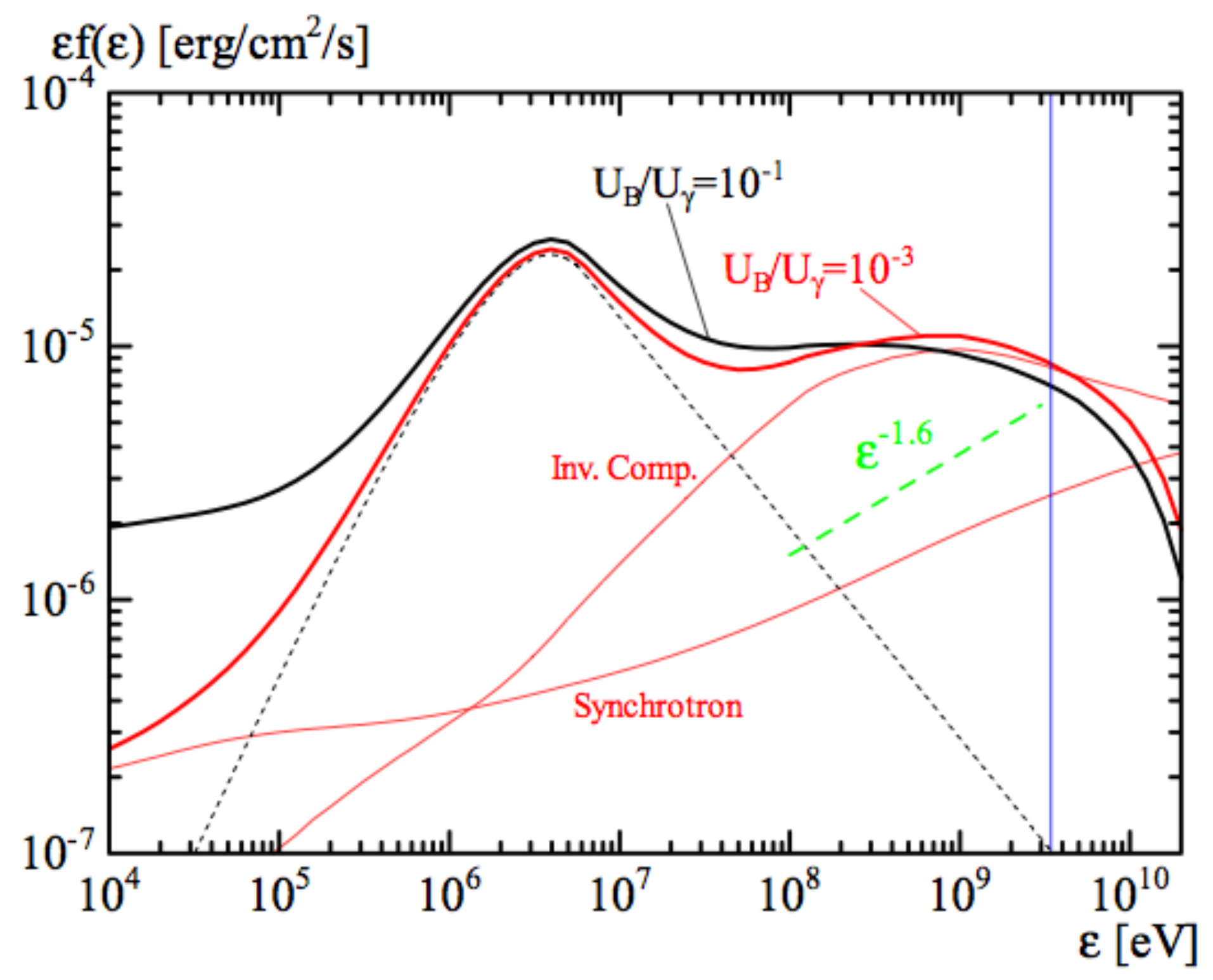}
\caption{ 
Photohadronic cascade radiation model for the additional power-law
component detected in GRB 090510. Synchrotron radiation from charged
pions, muons and $e^+e^-$ contribute to the observed spectra.
High-energy neutrinos are also expected in this model.
From \cite{Asano+09}. }
\label{fig:pgamma}
\end{figure}

An alternate mechanism to produce $\gtrsim 100$~MeV radiation in the
prompt GRB phase is through hadronic emission, either
proton-synchrotron radiation and radiation from $\gamma\gamma\to
e^+e^-$ pair cascades~\citep{Razzaque+10}
(Figure \ref{fig:psynchrotron}) or photohadronic interactions and
associated cascade radiation~\citep{Asano+09, Murase+12}
(Figure \ref{fig:pgamma}).  In the one-zone radiating shells, the
hadronic models can account for the delayed detection of the hard
power-law component through required proton acceleration and cooling
time.

The total energy of the burst, however, needs to be rather large in
hadronic models because of a high jet Lorentz factor, $\Gamma \gtrsim
500$, required to avoid $\gamma\gamma$ pair production {\it in-situ}
by $\gtrsim 1$~GeV photons.  In particular, to generate the observed
$2.44$~s delay in LAT for GRB~110731A the comoving frame jet magnetic
field needs to be $B^\prime \approx 10^5 (\Gamma/500)^{-1/3} (t_{\rm
onset}/2.44~{\rm s})^{-2/3}$ $(E_\gamma/100~{\rm MeV})^{-1/3}$~G, in
case of proton-synchrotron model. The corresponding
isotropic-equivalent jet luminosity is $L_{\rm jet} \gtrsim 10^{57}
(\Gamma/500)^{16/3} (t_{\rm onset}/2.44~{\rm s})^{2/3}$
$(E_\gamma/100~{\rm MeV})^{-2/3}$~erg~s$^{-1}$ \cite{Razzaque+10,
Wang+09}.  While the hadronic models require a larger total energy
than the leptonic models, the energy budget can be brought down for
$\Gamma < 500$, {see also \cite{Asano+10, Crumley+12}.}

A high $\Gamma$ is not a prerequisite to avoid $\gamma\gamma$
absorption in case of two-zone radiation, in either leptonic or
hadronic models, where LAT emission comes from late internal
shocks~\cite{Li10, Razzaque+10, Zou+11, Toma+11}.  The late
internal shocks, however, can not explain temporally extended LAT
emission which is apparently non-variabile.

\begin{figure}
\includegraphics[width=3.in]{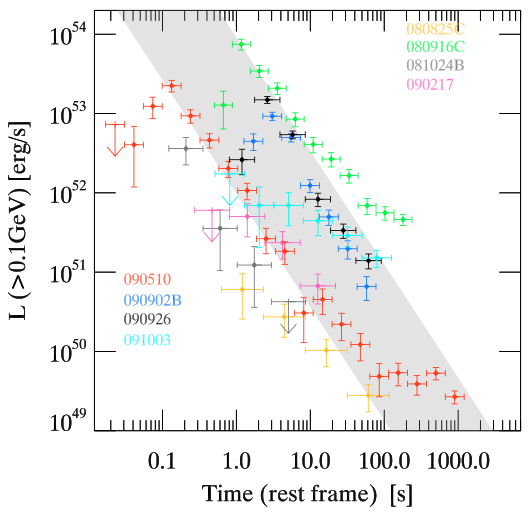}
\caption{ 
Light curves of $\gtrsim 100$~MeV radiation from a few LAT-detected
GRBs, which show decaying LAT flux as a common feature.  The grey band
with slope $t^{-10/7}$is the expected decay behavior for afterglow
emission from a radiative blast wave. From~\cite{Ghisellini+10}. }
\label{fig:LATflux}
\end{figure}

The delayed onset, longer duration and smooth temporal decay of the
$\gtrsim 100$~MeV flux (Figure \ref{fig:LATflux}) led to the
development of an early afterglow interpretation of high-energy
radiation from the external forward shock \cite{Kumar+10,
Ghisellini+10, DePasquale+10, Razzaque10, Corsi+10}.  The time delay
between keV--MeV radiation and $\gtrsim 100$~MeV radiation is
explained as the time required for deceleration of the GRB jet with
high $\Gamma$.  The hard power-law component and temporally-extended
LAT emission are interpretated as synchrotron radiation from the
external shock while keV--MeV emission from the internal shocks in
these two-zone emission models.  {Detailed modeling, however,
shows that an extrapolation of the external shock emission at earlier
time fails to reproduce all of the $\gtrsim 100$~MeV flux in the
prompt phase, requiring a significant contribution from the internal
dissipation regions \cite{ZhangBB+11, He+11, Liu+11, Maxham+11}.}

\begin{figure}
\includegraphics[width=3.2in]{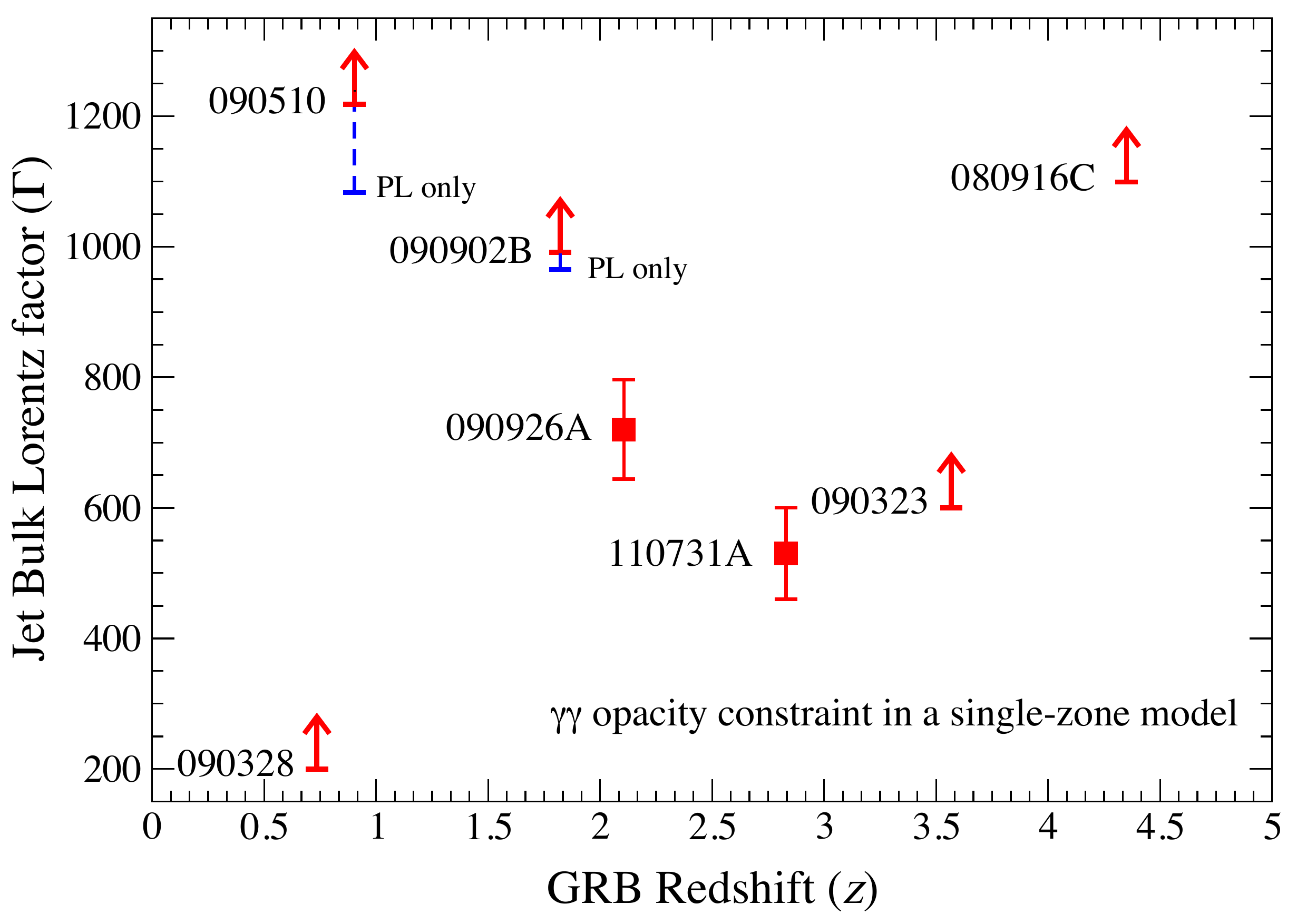}
\caption{
Bulk Lrentz factors of Fermi-LAT detected GRB jets calculated from
$\gamma\gamma\to e^+e^-$ pair absorption effect of GeV photons
assuming that they are produced in the same physical region as the
lower, $\lesssim 100$ MeV, photons.  The lower limits, $\Gamma_{min}$,
are calculated in cases where there were no cutoff in the spectra in
the GeV range.  In the ``PL only'' lower limits on $\Gamma_{min}$ were
calculated using only the additional polwer-law components of the
spectra. Estimates of $\Gamma$ rather than $\Gamma_{min}$ were made
when the observed $\gtrsim 1$ GeV cutoff in the spectra were asumed
due to $\gamma\gamma\to e^+e^-$ pair absorption.}
\label{fig:G0}
\end{figure}

Both in the internal- and external-shock scenario, modeling of
high-energy radiation by the leptonic and/or hadronic processes
depends crucially on $\Gamma$.  Currently $\gamma\gamma\to e^+e^-$
pair-production constraint in internal shocks is the most widely used
estimate of a lower limit on $\Gamma$~\citep{Krolik+91, Lithwick+01,
Razzaque+04}.  In the case of GRB~090926A and GRB~110731A a cut off in
the additional power-law component allowed to estimate $\Gamma$ rather
than its lower limit, assuming that the cut off at GeV energy is due
to attenuation of flux by $\gamma\gamma\to e^+e^-$ pair absorption.
Figure \ref{fig:G0} shows the estimates of $\Gamma$ from \fer
LAT-detected bright GRBs.  There are, however, large (a factor 2--3)
uncertainty due to theoretical modeling of target radiation fields and
environment~\citep{Baring+97, Granot+08, Ackermann+090510,
Hascoet+12}.

\subsection{GRB afterglow}
\label{subsec:afterglow}

Regardless of the details of the dissipation mechanism of the GRB jet
to produce prompt $\gamma$-ray emission, the accumulated surrounding
material by the jet drives a blast wave and external forward
shock \cite{Blandfor+76}. A reverse shock is also expected in the
baryon-loaded fireballs.  The deceleration takes place with the jet
energy compoarable to the kinetic energy of the swept-up material in
the blast wave, $E_{iso} = (4/3)\pi r_{dec}^3 n_{ext} m_p
c^2 \Gamma^2$. The deceleration radius is $r_{dec} \sim 2\times
10^{17} (E_{iso}/10^{53}~{\rm erg})^{1/2} (n_{ext}/1~{\rm
cm}^3)^{-1/2} (\Gamma/100)^{2/3}$ cm for a constant density external
medium.

After deceleration, the blast wave evolves in a self-similar fashion
with the bulk Lorentz factor decreasing as $\Gamma \sim t^{-3/8}$ for
an adiabatic fieball and $\Gamma \sim t^{-3/7}$ for a radiative
fireball, both in constant density medium \cite{Blandfor+76, Sari+98}.
Particles are expected to be accelerated in the forward shock, which
is highly relativistic.  Magnetic field also assumed to be generated
in the forward shock, which leads to synchrotron radiation by
relativistic particles \cite{Meszaros+97, Sari+98}.  As the jet
continues to be decelerated in the course of sweeping up more and more
external matter, the peak of the synchrotron flux shifts from
high-frequency to low-frequency bands.  Smooth decay of the flux
across different energy bands is a characteristic feature of the
afterglow synchrotron model \cite{Wijers+97}, whose predicted
detection allowed the first measurements of host galaxies and redshift
distances \cite{Costa+97, Paradijs+97}.

Broadband X-ray to radio emission from GRBs has been quite
successfully described by synchrotron radiation from a decelerating
blast wave with or without variation of the straightforward models.
Prompt optical emission \cite{Wijers+97}, at the time the deceleration
has been detected, with robotic ground-based telescopes such as ROTSE
and others. The external shock interpretation of the late
afterglow has proven robust overall, although some detailed features
such as the X-ray steep decays followed by flat plateaus and
occasional large flares are not well-understood yet.

\begin{figure}
\includegraphics[width=3.3in]{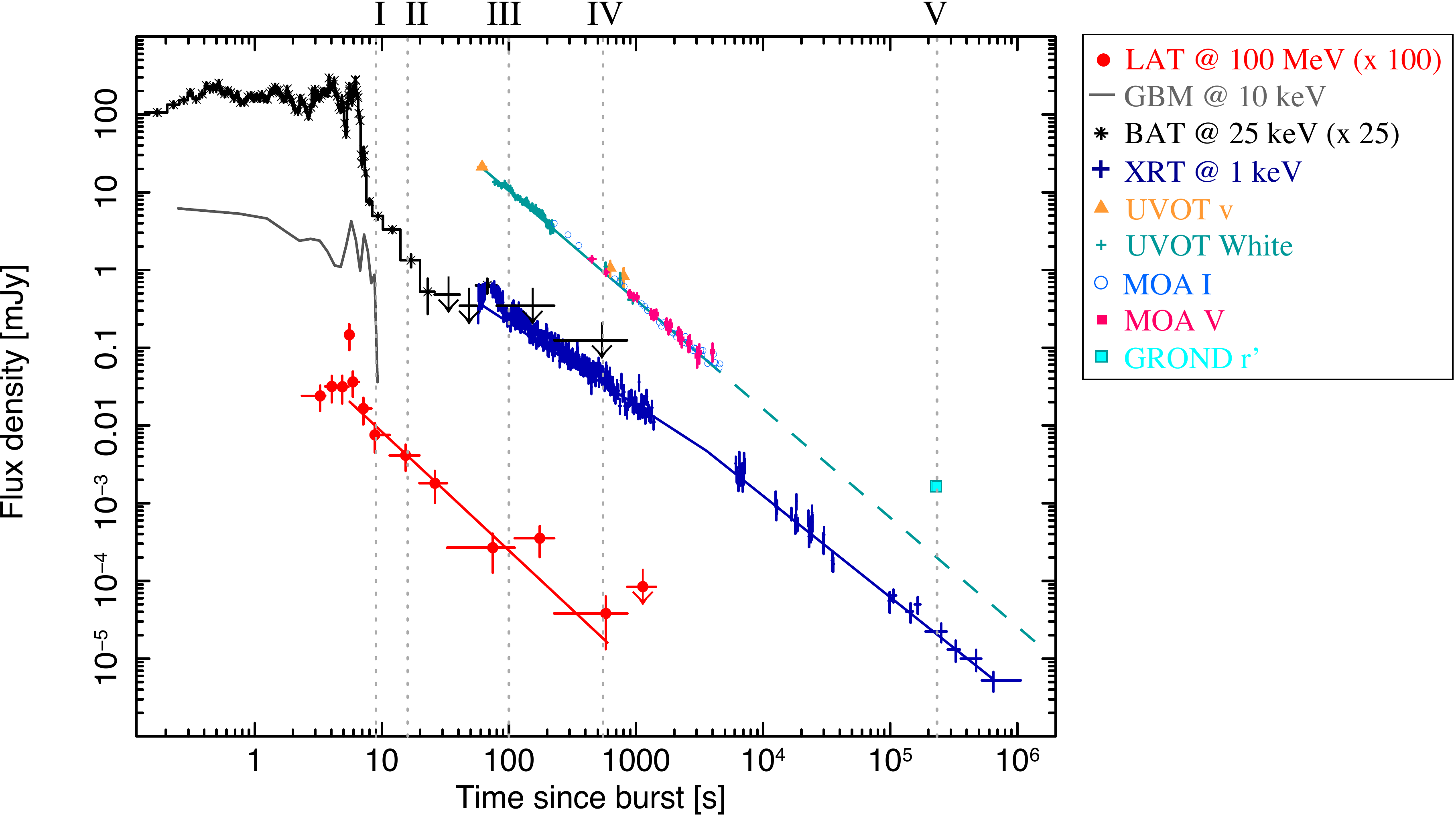}
\caption{ 
Multiwavelength light curves of GRB~110731A from {\em Fermi} GBM and
LAT; {\em Swift} BAT, XRT and UVOT; and ground-based optical
telescopes. Smooth decay of the fluxes across the wavelengths after
$\sim 6$~s suggests their origin, including LAT emission, as from a
decelerating blast wave in stellar wind
environment. From \cite{Ackermann+110731A}. }
\label{fig:GRB110731A_LC}
\end{figure}

\begin{figure}
\includegraphics[width=3.in]{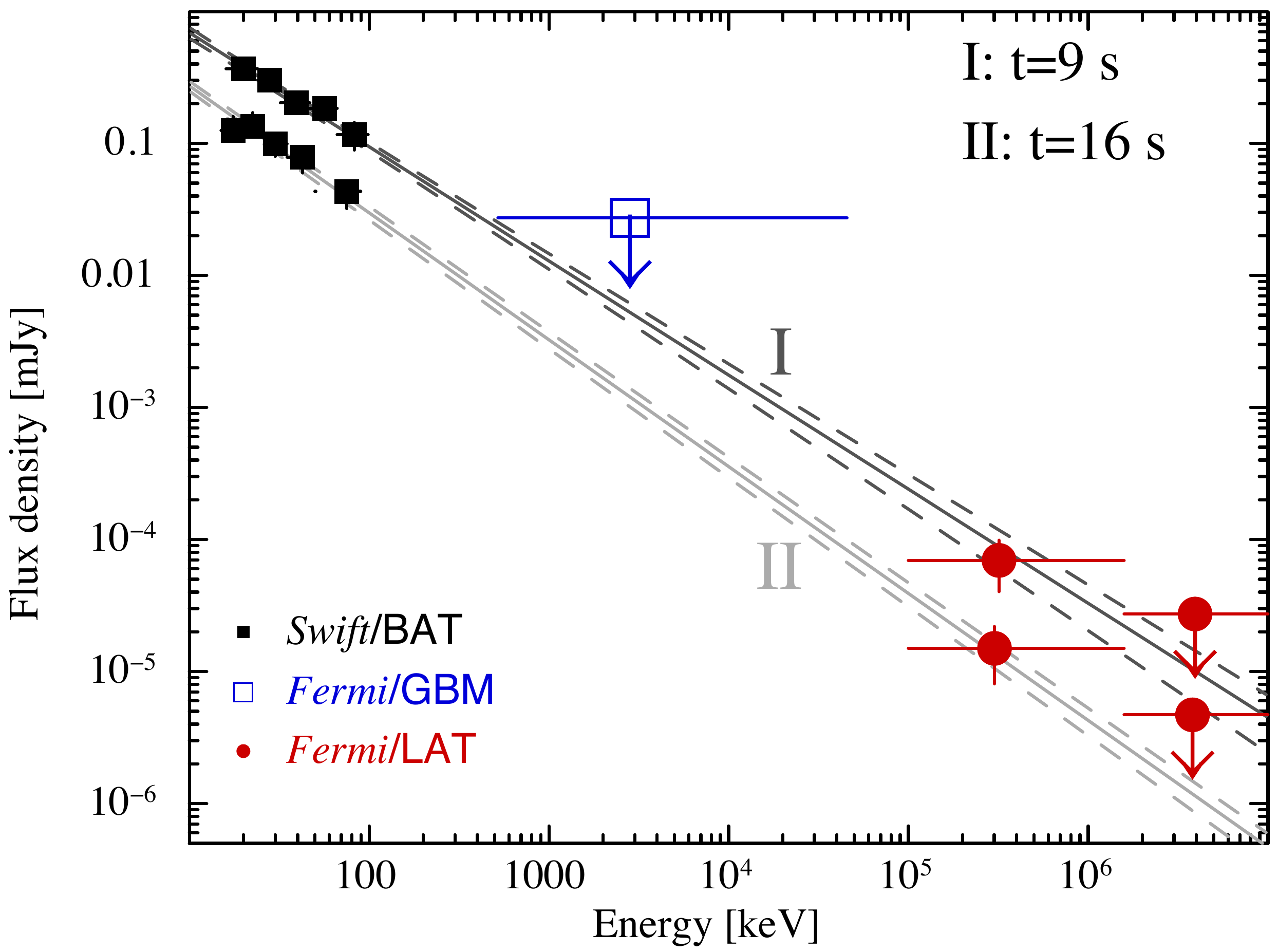}
\caption{ 
Broadband spectra of GRB~110731A at two time intervals after the
trigger. Single power laws fitting the spectra, together with flux
decay behavior in Fig.~\ref{fig:GRB110731A_LC} imply that LAT
emission is afterglow.  From \cite{Ackermann+110731A}.  }
\label{fig:GRB110731A_SED}
\end{figure}

As discussed in Section \ref{subsec:gev}, temporally extended $\gtrsim
100$ MeV emission detected by Fermi-LAT from many GRBs well after the
prompt keV--MeV emission has been advocated as afterglow emission by a
number of authors \citep{Kumar+10, Ghisellini+10, DePasquale+10,
Razzaque10, Corsi+10}.  To test an early afterglow interpretation of
the LAT data from GRBs one needs to confront models with simultaneous
multiwavelength data sets. Most recently, modeling of multiwavelength
light curves (Figure \ref{fig:GRB110731A_LC}) and broadband Spectral
Energy Distribution (SED) from GRB~110731A
(Figure \ref{fig:GRB110731A_SED}) strongly indicate
that \cite{Ackermann+110731A} (i) LAT extended emission as early as
$T_0+8.3$~s, from broadband SED, is compatible with afterglow
synchrotron emission in other bands. (ii) A wind afterglow
model \cite{Chevalier+00, Panaitescu+00b, Granot+02} is favored over
an ISM model, from the behavior of the light curves and SEDs. Detailed
modeling of GRB~110731A data also resulted in {$\Gamma \sim 500$}
just before deceleration of the blast wave \cite{Ackermann+110731A}.
An internal shock contribution to LAT data before $T_0+8.3$~s can not
be ruled out, however, and requires detailed investigation with future
Swift-Fermi bursts.

\subsection{High-energy neutrinos from GRBs}
\label{subsec:neutrino}

Neutrinos with energies $\gtrsim$ 100 TeV are expected from
photohadronic ($p\gamma$) interactions of relativistic protons
accelerated in the fireball internal or external shocks.  In the case
of internal shocks, the interaction is expected to take place between
MeV photons produced by radiation from accelerated electrons and
protons which are co-accelerated in the same shocks \cite{Waxman+97,
Rachen+98, Dermer+03, Guetta+03}. Neutrinos are expected to be
produced more efficiently at an energy $\varepsilon_\nu \gtrsim 100
(\varepsilon_{\gamma,pk}/100~{\rm keV})^{-1} (\Gamma/316)^2$ TeV.
Neutrino production in internal shocks is expected to suppress at an
energy $\epsilon_\nu \gtrsim 1$ PeV due to synchrotron and adiabatic
losses of the high-energy pions and muon \cite{Rachen+98, Kashti+05,
Lipari+07}.  In the case of external shocks, optical/UV photons
predominantly take part in $p\gamma$ interactions to produce $\gtrsim
1$ EeV neutrinos \cite{Vietri98, Waxman+99, Dai+01}.

\begin{figure}
\includegraphics[width=3.3in]{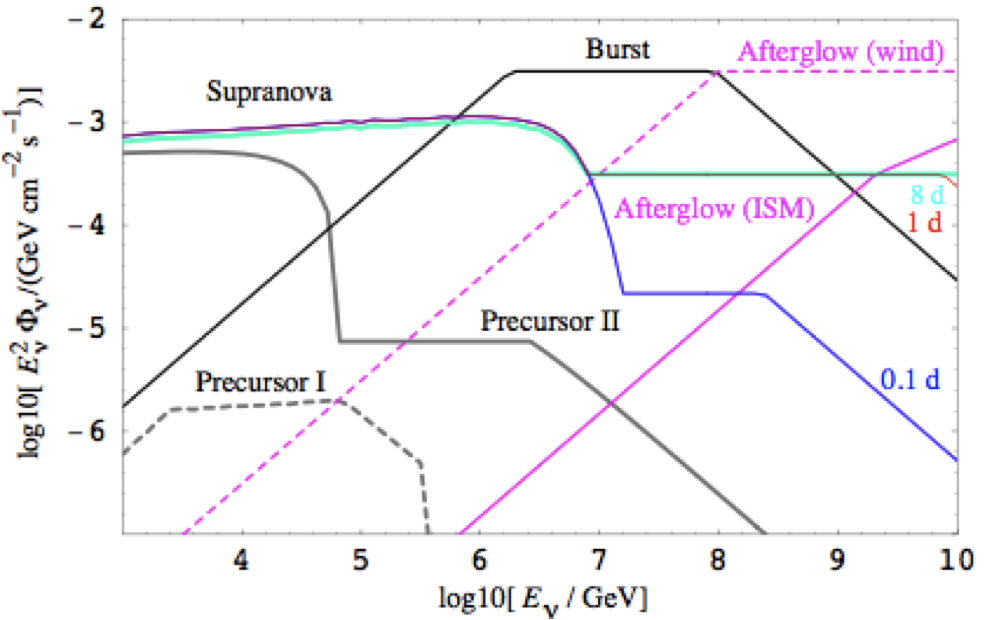}
\caption{ 
Predicted neutrino fluxes from a relatively nearby GRB 030329 at a
redshift $z=0.17$. An accompanying supernova was detected from this
long GRB as well. From \cite{Razzaque+04b}.}
\label{fig:nuflux}
\end{figure}

The most intense neutrino emission at $\lesssim 10$ TeV energies from
GRBs or collapsars might be due to $p,\gamma$ and $pp$ interactions
occurring {\it inside} the stellar envelope while the jet is still
burrowing its way out of the star \cite{Meszaros+01, Razzaque+03} and
before it successfully break through the stellar envelope to produce a
GRB or choke inside.  Figure \ref{fig:nuflux} shows different neutrino
flux models \cite{Razzaque+04b} from a nearby GRB 030329 at $z=0.17$
which was also connected with a supernova.

The IceCube Collaboration has recently constrained high-energy
neutrino emission from the internal shocks by stacking electromagnetic
data from GRBs \cite{Abbasi+11, Ahlers+11, Abbasi+12}.  This
constraint, in the $\sim 1$--20 PeV energy range, calculated from 40
and 59 string data severely excludes efficient neutrino production in
the internal shocks.  This could be due to the high bulk Lorentz
factor, $\gtrsim 500$, of the GRB jet which reduces photopion
production efficiency, {e.g.\ \cite{Razzaque+09, Crumley+12},} or
due to inefficient aceleration of protons in the internal
shocks.  {To accommodate IceCube non-detection, revisions to the
neutrino flux from internal shocks have also been
proposed \cite{Hummer+11, He+12}.  New calculations in the models
alternate to internal shocks are being done as well,
e.g.\ \cite{Zhang+12}.}  Data from the full IceCube detector will
either constrain the neutrino production efficiency in the internal
shocks or rule out these models.  Note that TeV neutrinos from the
buried GRB jet \cite{Meszaros+01, Razzaque+03} and EeV neutrinos from
the afterglow \cite{Vietri98, Waxman+99, Dai+01} remain viable.

\subsection{Gravitational waves from GRB}
\label{subsec:gw}

Binary mergers of NS-NS and/or NS-BH, the most popular model for short
GRBs, are the strongest candidate sources of gravitational waves (GWs)
with a total energy reaching $\sim 10^{54}$ ergs \cite{Kochanek+93,
Ruffert+99}.  The luminosity of GW from long GRBs, originating from
core-collapse of massive stars, is highly
model-dependent \cite{Fryer+01}.  Recent numerical calculations of GW
emission from long GRBs range from pessimistic \cite{Ott+11} to
modest \cite{Kiuchi+11} predictions.  The expected detection rates of
GW from compact binary mergers in coincidences with short GRBs have
been estimated to be several per year \cite{Abadie+10} for the
advanced LIGO and VIRGO. Coordinated detection of GW with
electromagnetic signal helps to reduce background significantly, if
short GRBs are the sources. {Combining sub-threshold signal from
electromagnetic, neutrino and GW observatories can also lead to
significant detection \cite{Smith+12}.}

\section{Future GRB observations}
\label{subsec:future}

Both Swift and Fermi are healthy and adequately-funded by NASA and
foreign partners.  Neither has lifetime limits from expendable
propulsion gasses or cryogens, and both have orbital lifetimes beyond
2025.  They are joined by INTEGRAL (e.g., Mereghetti et
al. 2003 \cite{Mereghetti+03}) and MAXI \cite{Serino+10}, which also
detect GRBs and the InterPlanetary Network of GRB
detectors \cite{Hurley+12} that localize events by triangulating their
arrival times.

There are several future missions with GRB capabilities that are being
developed or proposed.  They include:

\begin{enumerate}
\item 
SVOM -- A Chinese-French mission to detect GRBs and perform rapid
optical and X-ray follow-up \cite{Paul+11}.
\item 
UFFO -- A Korean payload flying on the Russian Lomonosov mission
\cite{Jung+11}.  The payload includes a wide-field hard X-ray burst detector
and a rapidly-orienting optical telescope.  The pathfinder version of
the mission will fly in 2012.
\item 
Lobster -- A wide-field focusing X-ray instrument operating in the
0.4--4 keV band.  There are several versions that have been proposed
including a UK instrument on the International Space Station \cite{Fraser+02}, a
NASA Explorer mission including an afterglow IR telescope \cite{Gerhrels+12}, and
an Explorer Mission of Opportunity on the International Space Station
\cite{Camp+12}.
\item 
LOFT -- A large-area X-ray timing mission selected for a Phase A study
for ESA's M2 opportunity \cite{Feroci+10}.
\item 
ASTROSAT -- An Indian mission with hard X-ray imager, scanning sky
monitoring, X-ray and UV telescopes \cite{Manchanda+11}.
\end{enumerate}

Fermi LAT detection of $> 10$~GeV photons indicate that the prospects
for detection of VHE radiation from GRBs by ground-based telescopes
are promising.  Highly sensitive Imaging Atmospheric Cherenkov
Teleescopes (IACTs) such as MAGIC, VERITAS and HESS with a threshold
energy as low as 25~GeV (special GRB mode for MAGIC) are currently
slewing to the GRB directions.  {Water Cherenkov detector HAWC is
also under construction.}  Simultaneous observations by Fermi LAT and
Cherenkov telescopes in the overlapping energy range will be extremely
helpful to understand the nature of high-energy radiation from
GRBs. Such observations can also pose strong constraints on both the
leptonic and hadronic models, especially independent handles on the
magnetic field strength are possible.  Long-sought SSC radiation from
the external forward shock or hadronic radiation in the prompt phase
can also be searched for and/or constrained.  Detection of very-high
energy radiation from GRBs can be modeled to estimate the total
explosion energy, which has implications for Graviational Wave
detection by LIGO/VIRGO.

There is a great opportunity for new science coming from future
observations coordinated with GRBs of gravitational waves, neutrinos
and across the electromagnetic spectrum.  There are major new
facilities coming on line and planned such as LIGO/VIRGO, IceCube,
PanSTARRS, LSST and the 30m telescopes.

\section{Conclusions}

We are in an era of rapid progress in understanding GRBs.  Swift,
Fermi, INTEGRAL, MAXI, and IPN are all detecting GRBs.  A vibrant
community of observers is following up the GRB detections with new and
capable instruments across the electromagnetic spectrum.  Key findings
from the past several years has included the following:

\begin{enumerate}
\item X-ray afterglow features, steep and shallow decays, flares,
\item afterglows and accurate localizations of short bursts supporting a possible merger origin,
\item confirmation of long GRB-Supernova connection,
\item detection of high-redshift, $z>5$, GRBs,
\item delayed onset of GeV emission,
\item high bulk Lorentz factor of the jet.
\end{enumerate}

With all this progress, there are many objectives that we hope to
accomplish in the next 10 years.  These include:

\begin{enumerate}
\item confirmation of short burst origin,
\item jet opening angle and measurements of lighcurve breaks
\item metallicity dependence
\item relation to star formation history
\item nature and interpretation of high energy spectral components.  
\end{enumerate}

We look forward to many years more of GRB discovery.

\section{Acknowledgements} 

We thanks John Cannizzo {and Charles Dermer} for valuable
consultation {and comments} on this paper.  All data from the
Swift and Fermi observatories are publicly available through NASA's
High Energy Astrophysics Science Archive Research Center (HEASARC) at
http://heasarc.gsfc.nasa.gov/.  Work of S.R. was supported by NASA
Fermi Guest Investigator program and was performed at the Naval
Research Lab while under contract.

\end{document}